\documentclass[conference]{llncs}
\usepackage{cite}
\usepackage{amsmath,amssymb,amsfonts}
\usepackage{graphicx}
\usepackage{textcomp}
\usepackage{xcolor}
\usepackage{url}
\usepackage{diagbox}

\usepackage{subfigure}

\def\BibTeX{{\rm B\kern-.05em{\sc i\kern-.025em b}\kern-.08em
    T\kern-.1667em\lower.7ex\hbox{E}\kern-.125emX}}
\begin{document}

\title{Survey and Analysis of DNS Filtering Components}

\author{Jonathan Magnusson}
\authorrunning{J. Magnusson}
\institute{Karlstad University\\Karlstad, Sweden\\
\email{jonathan.magnusson@kau.se}}

\maketitle

\begin{abstract}
    The Domain Name System (DNS) comprises name servers translating domain
    names into, commonly, IP addresses. \emph{Authoritative} name servers hosts
    the resource records (RR) for certain zones, and \emph{resolver} name
    servers are responsible for querying and answering DNS queries on behalf of
    their clients.  Unfortunately, cybercriminals often use DNS for malicious
    purposes, such as phishing, malware distribution, and botnet communication.
    To combat these threats, filtering resolvers have become increasingly
    popular, employing various techniques to identify and block malicious
    requests. In this paper, we survey several techniques to implement and
    enhance the capabilities of filtering resolvers including response policy
    zones, threat intelligence feeds, and detection of algorithmically
    generated domains.  We identify the current trends of each area and find
    missing intersections in the literature, which could be used to improve the
    effectiveness of filtering resolvers.  In addition, we propose future work
    designing a framework for filtering resolvers using state-of-the-art
    approaches identified in this study. 
\end{abstract}

\keywords{literature study \and
domain name system \and
response policy zone \and
threat intelligence feed \and
domain generation algorithm \and
resolver \and
filtering resolver \and
DNS firewall}

\section{Introduction}
Domain Name System (DNS) has become a crucial infrastructure component in the
modern Internet era. DNS is responsible for converting user-friendly domain
names into their corresponding IP addresses. The DNS resolvers act as a
mediator between the end-users and the authoritative name servers, and they
play a critical role in ensuring the smooth functioning of the Internet.
However, with the increasing prevalence of cyber threats, the need for secure
and efficient DNS resolvers has become more critical. One way to enhance the
security of DNS resolvers is through filtering.  DNS filtering blocks or allows
specific queries and responses based on predefined rules similar to a
firewall.  This technique prevents users from accessing malicious or
inappropriate content, enforces organizational policies, and protects against
cyber threats such as phishing, malware, and botnets.  By implementing DNS
filtering, network administrators can prevent employees or guests from
accessing potentially harmful websites or applications, thereby improving the
organization's overall security posture.

This study aims to lay the groundwork for a future framework incorporating
state-of-the-art approaches to filtering in DNS resolvers.  Surveying the
literature around three techniques for identifying and mitigating malicious DNS
traffic, we analyze trends and highlight the advantages and limitations of each
technique, the effectiveness of each approach in preventing cyber threats, and
the challenges associated with their implementation.  The three techniques
studied in this survey are: response policy zone (RPZ), threat intelligence
feed (TIF), and domain generation algorithm (DGA) detection.

The paper is structured as follows: 
Section~\ref{sec:bg} introduce the relevant background for the study and
highlights the lack of literature surveys in the area.
Section~\ref{sec:method} presents the methodology behind the literature survey.
Section~\ref{sec:litstu} summarizes the resulting papers. In
Section~\ref{sec:disc} the results are discussed, including trends within each
topic and the intersection between them, identifying areas for future research.
Section~\ref{sec:con} summarize and concludes the study.

\section{Background} \label{sec:bg}
In this section we present the various areas related to this study, including
name servers, response policy zones, threat intelligence feeds, malware,
command-and-control servers, and domain generation algorithms as well as how
they relate to each other. 

\subsection{Name Servers}
DNS is a widely used hierarchical key/value store that facilitates the mapping
of domain names to resource records (RRs) on the Internet~\cite{RFC1034,
RFC1035}. One of the primary functions of DNS is to perform lookups for A/AAAA
RRs containing IPv4 and IPv6 addresses, respectively.  The RRs are located in
zones, which are portions of the DNS namespace operated by a distinct
organizations or administrators.  A name server in DNS can be
\emph{authoritative} for a zone and keep track of RRs, or it can be a
\emph{resolver} that provides answers to queries from a cache or by querying
the authoritative name servers.  Each client has a \emph{stub} resolver that
listens to API calls and sends a query to a preconfigured resolver. This
preconfigured resolver can either be a \emph{recursive} resolver performing
lookups against the authoritative name servers or a \emph{forwarding} resolver
sending the query to another resolver.  This configuration results in a
resolver chain that begins with the stub resolver and ends with the recursive
resolver, with any number of forwarding resolvers in between them. The answer
to a query may be cached at any point in the resolver chain to improve
performance.  A resolver can be configured to serve a specific set of clients,
such as a local resolver for a private network or an ISP resolver for clients
connected to the network. A public resolver such as Google Public
DNS~\cite{googledns} and Cloudflare DNS~\cite{cloudflare} is open to anyone.

\subsection{Response Policy Zones}
\begin{figure}[h]
\centering
\includegraphics[scale=0.7]{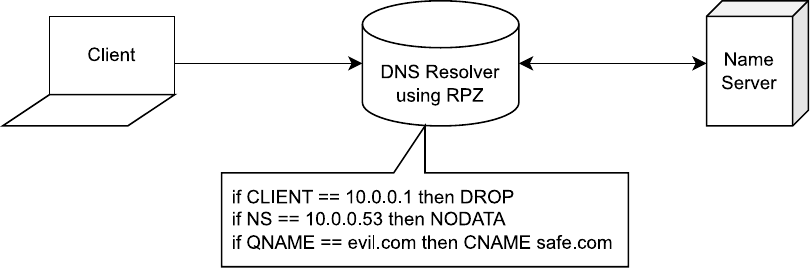}
\caption{Resolver using a response policy zone.}
\label{fig:rpz}
\end{figure}

Response policy zones (RPZs) is a feature in resolvers allowing administrators
to specify policies for DNS responses based on domain names and
context~\cite{rpz}. This reputation-based DNS firewall technology works by
intercepting DNS queries and responses, and checking them against a list of
policies defined by the administrator.  When finding a policy match, the
resolver returns a response according to the policy, such as a ``blocked'' or
``redirected'' response (see Figure~\ref{fig:rpz}). 

RPZs are a powerful tool for DNS administrators to enforce security policies
and protect their networks from malicious activities.  They offer flexibility
and customization, allowing administrators to create policies tailored to their
needs. However, properly configuring and maintaining RPZs are critical to avoid
false positives and maintain proper policy enforcement.  The policies need to
be kept up-to-date with accurate information, and can only protect against
known threats.

\subsection{Threat Intelligence Feeds}
\begin{figure}
  \centering
  \subfigure[Clients sharing traffic data.]{
      \label{fig:tif1}
      \includegraphics[width=0.45\textwidth]{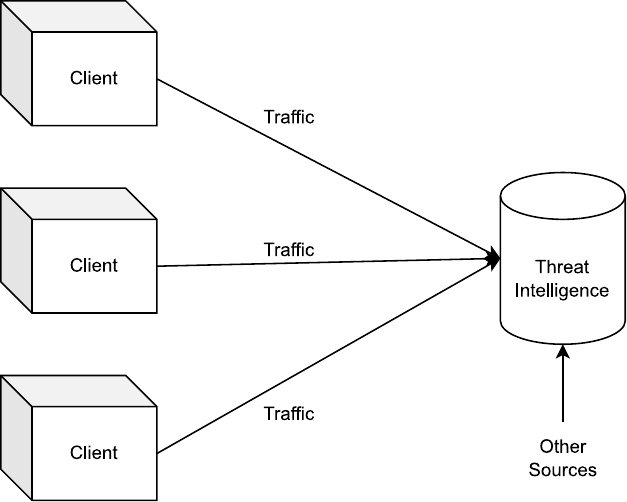}}\qquad
  \subfigure[Threat intelligence sharing results.]{
      \label{fig:tif2}
      \includegraphics[width=0.45\textwidth]{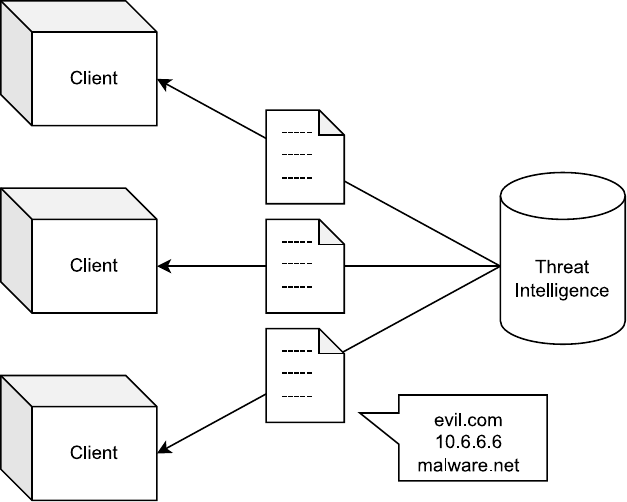}}
\caption{Threat intelligence feed system.}
\label{fig:tif}
\end{figure}

Cyber threat intelligence refers to information about potential or actual cyber
threats that can be used to inform and improve an organization's security
posture~\cite{ti1, ti2, gao2021}.  This information can include details about the
tactics, techniques, and procedures (TTPs) used by known threat actors, as well
as indicators of compromise (IOCs) such as IP addresses, domain names, and
hashes of malware.  In the context of DNS, threat intelligence feeds (TIFs) 
provide up-to-date information about known malicious domains
and related IP addresses (see Figure~\ref{fig:tif}).  

TIFs can be obtained from various sources, including open-source intelligence,
commercial vendors, and industry-specific threat-sharing groups. These feeds
contain a list of IOCs typically generated through analysis of network traffic,
security logs or other threat intelligence sources. The feeds are continuously
updated to reflect new threats emerging and can be customized to focus on
specific types of threats or geographies.  Filtering resolvers can use these
threat intelligence feeds to block or redirect DNS queries to known malicious
domains.  Using threat intelligence feeds in DNS filtering provides a powerful
defense against known threats and can help organizations detect and prevent
attacks before they can cause damage. However, it is essential to remember that
TIFs are not silver bullets and must be used in conjunction with other security
measures to provide comprehensive protection. Additionally, \emph{false
positives} can occur, leading to the blocking of legitimate domains.
Therefore, careful tuning and monitoring of feeds are necessary to ensure
effective use.

\subsection{Malware and Command-and-Control severs}
Malware is malicious software that infects computers to carry out a variety of
harmful actions, such as stealing sensitive information, causing damage to the
system, or using the device to launch attacks on other systems~\cite{malware}.
Botnets are collections of compromised devices controlled by a single entity
called the ``botmaster''~\cite{botnets}. Botnets often carry out large-scale
attacks, such as distributed denial-of-service (DDoS) attacks or spam
campaigns.  A device infected with malware will typically communicate with a
Command-and-Control (C2) server to receive instructions on what actions to
carry out. C2 servers are critical components of malware, and by blocking
traffic to and from them, organizations can reduce the damage from infected
devices. 

\subsection{Domain Generation Algorithms}
\begin{figure}[h]
\centering
\includegraphics[scale=0.6]{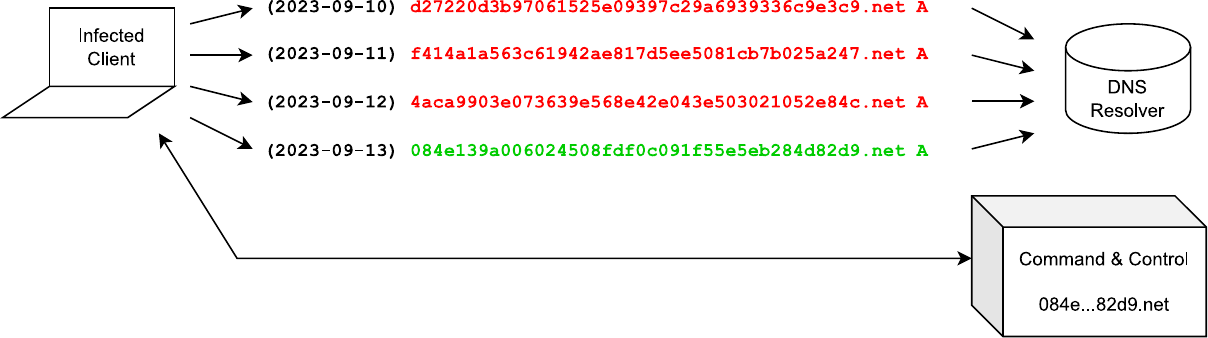}
\caption{Infected client generating domains based on current date,
    eventually finding the C2.}
\label{fig:dga}
\end{figure}

Domain generation algorithms (DGAs) is a technique used by malware to generate
unique domain names used for C2 communication~\cite{knysz2011, stone2009}.
DGAs use algorithms to generate unique domain names, generally using the
current date and time as input, making it difficult for filtering systems to
detect and block them in time if they rely on static or slow blocklisting of
known malicious domains or IP addresses. Malware using a DGA generates new
domain names periodically and attempt to establish communication with the C2
server (see Figure~\ref{fig:dga}).  
The botmaster may register only \emph{one} predetermined domain name for the C2
server in order to communicate with the infected devices. They can, however,
be detected with machine learning (ML) algorithms~\cite{mitchell1997}
identifying underlying patterns in algorithmically generated domain names. By
analyzing the structure of malware-generated domain names as well as query
behavior and context, it is possible to detect and block communications with
the C2 server, even if the domain names only appear once.

\section{Methodology} \label{sec:method}
RPZ, TIF and DGA detection have garnered particular attention from researchers
and practitioners in recent years, exploring their merits and limitations.
However, it is essential to note that despite these valuable contributions
(presented in Section~\ref{sec:litstu}), a literature survey still needs to
synthesize and evaluate the broader landscape of DNS resolver filtering and how
these techniques fit together.  The lack of existing literature surveys
covering DNS resolver filtering techniques underscores the need for a thorough
investigation into their roles, potential advantages, and any limitations they
may have in enhancing DNS security. This study adopts a \emph{specialized}
approach rather than providing a comprehensive overview of all aspects of DNS
resolver filtering; by doing so, it aims to provide a resource that offers
practical guidance and a consolidated state-of-the-art view of these
techniques.

While focusing intensely on specific techniques, as exemplified by this study,
allows for in-depth analysis and the generation of practical insights, it may
necessitate overlooking broader aspects of DNS resolver filtering.
Comprehensive surveys strive to encompass the entire landscape, offering a
holistic understanding of the field.  However, as demonstrated in this study,
the decision to concentrate on specific techniques is deliberate, driven by the
need to investigate the intricacies and nuances of these specialized methods
thoroughly. By doing so, we aim to offer practical insights and nuanced
comprehension of their role and potential advantages within the broader DNS
resolver filtering ecosystem. We encourage future research to explore the
broader landscape of DNS resolver filtering, encompassing additional techniques
and strategies, advancing our collective understanding of this critical field.

This literature survey comprise a review of academic research papers related to
filtering resolvers, focusing on the use of response policy zones, threat
intelligence feeds, and domain generation algorithms.  The literature search
used three academic databases: 
Google Scholar\footnote{\url{https://scholar.google.com/}}, 
ACM Digital Library\footnote{\url{https://dl.acm.org/}}, and 
IEEE Xplore\footnote{\url{https://ieeexplore.ieee.org/}}.  
The search terms used were:

\begin{itemize}
    \item \textbf{RPZ}: "RPZ" OR "Response Policy Zone"
    \item \textbf{TIF}: "TIF" OR "Threat Intelligence Feed"
    \item \textbf{DGA}: "DGA" OR "Domain Generation Algorithm"
\end{itemize}

\begin{table}[h!]
    \begin{center}
    \caption{Database search with topics}
    \label{table:dbkey}
    \begin{tabular}{|l|c|c|c|}
        \hline 
        \diagbox[]{Database}{Topic} & RPZ & TIF & DGA \\\hline
        \hline
        Google Scholar & 109 & 60 & 33 \\\hline
        ACM Digital Library & 4 & 5 & 5 \\\hline
        IEEE Xplore & 3 & 0 & 0 \\\hline
    \end{tabular}
    \end{center}
\end{table}

To find relevant results we add ``DNS'' to each search.  Initial results of DGA
yielded over 3000 results. To narrow down our search we anchored DGA to either
RPZ or TIF: \texttt{DGA AND (RPZ OR TIF)}.  The search was limited to articles
published in English from 2018 to 2023, excluding technical reports and thesis
work. The number of results are seen in Table~\ref{table:dbkey}.  Duplicate results
were then removed within each category and titles were screened for relevancy,
resulting in 14 RPZ, 12 TIF and 13 DGA.  None of the papers showed up in all
three categories, 13 papers showed up in two categories and 13 papers only
showed up in one category. The papers were then screened by abstract and
labeled as: \{INCLUDE, EXCLUDE\}. (see Table~\ref{table:papertopic}) The papers
were screened by intro and conclusion if abstract was not enough. The screening
resulted in 22 papers.

\begin{table}[h!]
    \begin{center}
        \caption{Papers, topics and status (sorted by publication year)}
    \label{table:papertopic}
    \begin{tabular}{|l|c|c|c|c|}
        \hline 
        Paper & RPZ & TIF & DGA & Status\\\hline
        \hline 
        Arnaldo et al. 2018 & & X & X & IN\\\hline
        Chin et al. 2018 & & X & X & IN\\\hline
        Najafi et al. 2018 & & X & X & EX\\\hline
        Zrahia et al. 2018 & & X &  & EX\\\hline
        Rweyemamu et al. 2019 & & X &  & IN\\\hline
        Li et al. 2019a & & X &  & IN\\\hline
        Li et al. 2019b & & X & X & IN\\\hline
        Satoh et al. 2019 & X & & X & IN\\\hline
        Spacek et al. 2019 & X & &  & IN\\\hline
        Wilde et al. 2019 & X & &  & IN\\\hline
        Griffioen et al. 2020 & & X & X & IN\\\hline
        Jin et al. 2020 & X & &  & IN\\\hline
        Magalhaes et al. 2020 & X & & X & IN\\\hline
        Noor et al. 2020 & X & &  & EX\\\hline
        Drichel et al. 2021 & & X & X & IN\\\hline
        Fejrskov et al. 2021 & X & &  & IN\\\hline
        Jin et al. 2021 & X & & X & IN\\\hline
        Satoh et al. 2021 & X & & X & IN\\\hline
        Drichel et al. 2022 & & X & X & IN\\\hline
        Ichise et al. 2022 & X & &  & IN\\\hline
        Mohus et al. 2022 & X & &  & IN\\\hline
        Preuveneers et al. 2022 & & X &  & IN\\\hline
        Villalon-Huerta et al. 2022 & & X &  & EX\\\hline
        Wala et al.	2022 & X & & X & IN\\\hline
        Jin et al. 2023 & X & &  & IN\\\hline
        Mitsuhashi et al. 2023 & X & & X & IN\\\hline
    \end{tabular}
    \end{center}
\end{table}

\section{Surveyed Literature} \label{sec:litstu}
This section contains summaries of each paper in this literature survey. The
summaries are divided into the three topics based on primary contributions if
covered by two topics. 
Section~\ref{sec:lit:rpz} focuses on RPZ, where papers are summarized and
organized chronologically according to their publication year. This section
aims to consolidate and present an overview of the primary contributions made
within RPZ.  
Section~\ref{sec:lit:tif} summarizes the literature on TIF, using the same 
approach to summarizing and arranging papers. This section aims to elucidate
the primary contributions made in TIF.  
In Section~\ref{sec:lit:dga}, the literature primarily regarding DGA is
summarized and organized. This section aims to provide a comprehensive summary
of the primary contributions made in the context of DGA.

\subsection{Response Polizy Zone} \label{sec:lit:rpz}
\paragraph{Satoh et al. 2019} propose a cause-based classification approach for
identifying the causes of malicious DNS queries detected through
blocklists~\cite{satoh2019}.  The approach aims to reduce the number of
investigated malicious queries by limiting the investigation scope to only
\emph{representative} queries in the classification results. Through
experiments, the proposed approach grouped 388 malicious queries into three
clusters, each comprising queries with a common cause. The results show that
administrators could briefly pursue all the causes by investigating only
representative queries of each cluster, thereby addressing the problem of
infected machines in the network. The study's contribution shows that
surrounding queries can help classify malicious queries by their causes and the
potential of the approach for enhancing the detection and precise
classification of various malicious activities in multiple system logs through
the security information and event management system.

\paragraph{Spacek et al. 2019} examine the use of a DNS firewall as a
cybersecurity tool to filter access from the protected network to known
malicious domains outside the network~\cite{spacek2019}.  The authors
formulate functional requirements for a DNS firewall to fulfill the role of a
cybersecurity tool and developed it using the DNS RPZ technology. The authors
find limitations of RPZ during testing and propose possible solutions to make
the DNS firewall a more complex cybersecurity tool.  The DNS firewall is
suitable when used with other tools like IP blocking and automated malicious
domain detection, but it still has limitations in \emph{user informing} and 
\emph{blocklist sharing}. Despite these limitations, the authors prefer 
implementing the DNS firewall with RPZ and continue researching automation 
possibilities and firewall log visualization.

\paragraph{Wilde et al. 2019} discuss using RPZ to prevent malicious actors
from using DNS in Internet attacks~\cite{wilde2019}.  The authors compare the
blocking behavior of a freely available RPZ-enabled DNS service with other DNS
services available in the United States.  They find that only the RPZ-enabled
server was significantly blocking malicious domains. The authors suggest that
RPZ can provide an additional layer of defense against attacks that may bypass
browser-based defenses.  Organizations with many computer users should consider
adding RPZ to their defenses. The paper concludes that DNS firewalls using RPZ
have considerable potential to make the Internet safer and mitigate the current
imbalance between cyber-attack and cyber-defense.

\paragraph{Jin et al. 2020} propose a novel strategy for mitigating phishing
attacks through email-based URL distribution~\cite{jin2020}.  Instead of
investigating \emph{all} incoming emails, the proposed strategy involves
extracting URLs from incoming emails and registering the corresponding
fully-qualified domain name (FQDN) in the local DNS cache server's RPZ with
mapping to the IP address of a HTTP proxy. Traffic to the phishing URLs is
directed to the HTTP proxy, which is connected to different security facilities
conducting various inspections, thus avoiding heavy investigations on all
incoming emails and mitigating overloads on security facilities. The paper
evaluates the proposed strategy on a prototype system. It shows that URL
extraction and FQDN registration were completed before email delivery, and
traffic was successfully directed to the HTTP proxy. Overhead measurements
showed that the proposed strategy only affected the internal email server, 
with an 11\% decrease in performance on the prototype system.

\paragraph{Magalhaes et al. 2020} discuss the limitations of current DNS
firewall solutions that rely on blocklists to prevent users from accessing
malicious domains and proposes using ML to detect malicious domains in
real-time~\cite{magalhaes2020}.  The paper analyzes the viability of ML-based
detection using a large dataset of pre-classified malicious and benign domains.
Using the dataset enriched with multiple features, six ML algorithms classify
the domains, achieving accuracy rates between 75\% and 92\%. However, the
classification time ranged from 2.77 to 5320 seconds, indicating room for
improvement in accuracy and speed. The article emphasizes the importance of
detecting malicious domains on time, as malicious actors frequently register or
use previously unknown domains for short periods. The proposed ML-based
solution could provide real-time protection and block illegitimate
communications for previously unknown malicious domains. The article suggests
using the solution to implement a service through which users can submit
domains to be analyzed and receive an analysis report or provide enriched
datasets for the scientific community. The paper concludes that an ML-based
solution can contribute to developing a firewall solution with potential
application in a real context.

\paragraph{Fejrskov et al. 2021} propose using NetFlow data to measure the
effectiveness of DNS-based blocklists in preventing cybersecurity threats
instead of traditional methods that count the number of blocked DNS
requests~\cite{fejrskov2021}.  The study find that only a tiny percentage of
DNS responses match a blocklist entry, and a small proportion of these are
associated with an observed flow.  Additionally, many blocked flows are
\emph{potentially benign}, such as those toward a web server hosting both 
benign and malicious sites. The study highlights the importance of carefully 
choosing the type and category of blocklist before deployment to avoid 
undesired impact on users.

\paragraph{Jin et al. 2021} propose an anomaly detection system for detecting
and blocking malware-related outbound traffic on user terminals~\cite{jin2021}.
The system uses Software Defined Network (SDN) and RPZ technologies for
outbound traffic filtering. RPZ manages the registration of application
programs and DNS-resolved IP addresses, while SDN manages the network traffic
control. The prototype system is implemented on a MacOS machine and can be
deployed on a Windows system.  The system enables users to decide (allow or
deny) based on an alert message for detected outbound traffic to reduce false
positives.  This \emph{user informing} feature addresses one of the limitations
of using RPZ as a DNS firewall identified by Spacek~\emph{et
al.}~\cite{spacek2019}.

\paragraph{Ichise et al. 2022} address the security issue of bot-infected
computers sending direct outbound DNS queries to malicious DNS
servers~\cite{ichise2022}. The proposed solution is a policy-based detection
and blocking system using DNS RPZ. The authors implements a prototype system
and evaluate its performance in a local SDN-based network environment. The
preliminary evaluation show that the proposed system works as designed and
performs better than a previous system that used the MySQL database.

\paragraph{Wala et al. 2022} discuss the ``off-label'' use of DNS, where
legitimate entities use the DNS protocol for non-malicious purposes other than
domain resolution~\cite{wala2022}. The paper examines DNS logs collected over
a long period to reveal some of these use cases. It highlights how network
security defenders can leverage them to improve their detection on the network.
The paper also highlights the security implications of these off-label use
cases, including bypassing local DNS servers and firewall protection and the
abuse of local service provider networks and resources. Finally, the paper
suggests that with the advent of encryption in DNS, it will become harder to
investigate these off-label DNS use cases, and new ways to detect them under
encryption are actively being researched.

\paragraph{Jin et al. 2023} propose a mechanism for blocking access to
phishing URLs received through email by directing HTTP(S) access triggered by
clicking on the URLs to a particular proxy using DNS and RPZ
features~\cite{jin2023}. The system alerts users about the potential phishing
attack and allows them to either pass or block the access based on their
decisions. The proposed mechanism aims to protect end users from phishing
attacks while avoiding detailed investigations on all incoming emails. The
authors implement a prototype and conduct preliminary evaluations,
which confirm the system's functionality.

\subsection{Threat Intelligence Feeds} \label{sec:lit:tif}
\paragraph{Li et al. 2019a} aim to bridge the gap between available information
about up-to-date threats and the limited understanding of recipients in
effectively applying this data~\cite{li2019a}.  To achieve this, the
researchers define a set of metrics for characterizing TIFs (volume,
differential contribution, exclusive contribution, latency, accuracy, coverage)
and systematically apply them to a range of public and commercial sources.
External measurements are used to investigate issues of accuracy and coverage
qualitatively. The results suggest significant limitations and challenges when
using existing TIFs for their intended goals. The study reveals that feeds vary
significantly in the kinds of data they capture, with few explaining the
methodology behind the collection. Consumers are left with simple labels on
feeds such as "scan" or "botnet," while other labels like "malicious" and
"suspicious" are loosely defined, leaving interpretation up to consumers. The
study recommends standardization of data labeling to improve the effectiveness
of TIFs.

\paragraph{Griffioen et al. 2020} discuss the importance of cyber threat
intelligence in mounting an effective defense against
cyberattacks~\cite{griffioen2020}. It argues that specialized feeds of cyber
threat indicators can help organizations better understand their threat profile
and react on time to emerging threats. However, the effectiveness of these
feeds depends on their quality, since incorrect or incomplete information may
cause harm. The paper evaluates the quality of 24 open-source TIFs over several
months and finds significant variations in their performance. It also
highlights biases towards certain countries and potential collateral damage
when blocking listed IP addresses. Overall, the paper emphasizes the importance
of high-quality, timely, and relevant cyber threat intelligence in defending
against evolving threats.

\paragraph{Preuveneers et al. 2022} propose a solution for sharing cyber
threat intelligence between organizations while preserving the privacy of
sensitive information~\cite{preuveneers2022}.  Many organizations are reluctant
to share locally collected cyber threat intelligence due to the risk of
exposing sensitive business data or personally identifiable information. The
proposed solution is a practical polyglot approach that uses cryptographic
building blocks for private set intersection, and Bloom filters to efficiently
and effectively share threat intelligence. The paper also presents a novel
private graph intersection method that analyzes correlations between threat
events in a privacy-preserving manner and across sharing organizations. The
paper evaluates these techniques' security impact and computational overhead
and demonstrates the proposed solution's practical feasibility.

\subsection{Domain Generation Algorithm Detection} \label{sec:lit:dga}
\paragraph{Arnaldo et al. 2018} present an automated learning system that
leverages open-source data to develop a deep learning (DL) model capable of
detecting and publishing unreported malicious domains~\cite{arnaldo2018}.  The
system utilizes threat intelligence to label detected domains and updates its
detection models periodically. The authors highlight the problem of blocklists
from threat intelligence needing to be faster to update and lacking in
coverage.  Furthermore, the authors demonstrate that their system extends the
threat intelligence feed coverage and reduces detection delays. They also
present a DL model that can generate domains likely to be registered by
attackers in the future. 

\paragraph{Chin et al. 2018} propose a ML framework identifying and clustering
domain names for blocklists used to mitigate malware using DGA~\cite{chin2018}.
The researchers collects a real-time TIF of malicious
domains over six months and apply the proposed ML framework to study
DGA-based malware. The framework consists of a two-level model: classification
and clustering, which helps to identify DGA domains and the family/type of DGA.
The experimental results show a high accuracy of \textbf{95\%} for the 
classification and \textbf{92\%} for the clustering.

\paragraph{Rweyemamu et al. 2019} show that it is possible to infiltrate
domain rankings with trivial effort~\cite{rweyemamu2019}.  With just seven fake
users and 217 fake visits per day, it is possible to maintain a rank in Alexa's
top 100k domains\footnote{Discontinued on May 2022: \url{https://web.archive.org/web/20220102200605/https://support.alexa.com/hc/en-us/articles/4410503838999}}.  
Other studies rely on domain rankings when training ML models to classify
benign and malicious domains, assuming that popular domains are safe.  The
results suggest that future research should reconsider when using domain
rankings to model benign patterns. Discarding domains ranked for less than a
year is not enough to eliminate malicious domains. An attacker could easily
circumvent such measures by mounting a long-term attack on the list.  It is
therefore interesting and essential to detect rank manipulation in future
research.

\paragraph{Li et al. 2019b} propose a ML framework for detecting DGA
domains~\cite{li2019b}, comprising a dynamic blocklist, a feature extractor, a
two-level model, and a prediction model. The two-level model includes a
classification level and a clustering level. The prediction model uses a
time-series model based on a hidden Markov model (HMM) to predict incoming
domain features. The researchers collect data from a TIF over one year and
build a deep neural network (DNN) model to handle the large dataset and enhance
the proposed ML framework. Extensive experimental results demonstrate high
accuracy levels of \textbf{95\%} for the classification, \textbf{97\%} for the
DNN model, \textbf{92\%} for the clustering, and \textbf{95\%} for the HMM
prediction.

\paragraph{Drichel et al. 2021} describe EXPLAIN, a feature-based and
contextless DGA multiclass classifier~\cite{drichel2021}.  The authors compare
the performance of EXPLAIN with several state-of-the-art classifiers, including
DL models, and find that it achieves competitive results and is easier to
interpret than DL models. The predictions made by the proposed classifier can
be traced back to the characteristics of the used features. In contrast, DL
classifiers only output a vector of probabilities, indicating which class a
particular domain can be attributed to, without referring to the actual input.
The paper suggests that future work should compare the level of explainability
the proposed approach provides with different techniques which try to explain
the predictions of DNN classifiers. Their classifier achieved a F1-score of
\textbf{0.7855}, a precision of \textbf{0.8163}, and a recall of 
\textbf{0.7795}.

\paragraph{Satoh et al. 2021} propose an approach to identify and detect
malware with sophisticated DGAs that generate domain names dynamically by
concatenating words from dictionaries to evade detection~\cite{satoh2021}.  The
proposed approach analyzes the character strings of domain names at the word
level. It uses the distinct differences in word usage between malware-generated
and human-generated domains to identify malicious domains. The approach
achieves high accuracy, recall, and precision, effectively detecting dict-DGA
malware. Their classifier achieved an accuracy of \textbf{0.9989},
a recall of \textbf{0.9977}, and a precision of \textbf{0.9869}.

\paragraph{Drichel et al. 2022} present a study on detecting new
DGAs~\cite{drichel2022}. The current state-of-the-art classifiers are limited
in detecting new malware families as they can only assign domains to known
DGAs. The study evaluates four different approaches in 15 configurations. It
proposes a simple and effective approach that uses a classifier with a softmax
output layer and regular expressions to detect multiple unknown DGAs with high
probability while retaining state-of-the-art classification performance for
known DGAs. The evaluation is based on leave-one-group-out cross-validation
with 94 DGA families. The proposed approach, Regex Error Detection, performs
best, correctly attributing several unknown DGA classes with a F1-score of
\textbf{0.8011}, a precision of \textbf{0.8200}, and a recall of
\textbf{0.8063}. The classifiers are privacy-preserving, operating without
context and exclusively on a single domain name to be classified. The paper
also discusses class-incremental learning strategies that can adapt an existing
classifier to newly discovered classes and concludes that complete retraining
makes the most sense whenever a new DGA family is discovered.

\paragraph{Mohus et al. 2022} propose a novel method using a conditional GAN
(cGAN) to generate DNS data that could evade detection in an adversarial
setting~\cite{mohus2022}. The generator inputs random noise and static
fields necessary for the DNS data to work correctly. The discriminator works 
as a standard GAN, inputting samples from the GAN or training set,
including the output from other unsupervised detectors. During the training
phase, the cGAN backpropagates the discriminator loss to the discriminator
and the generator. This approach provides defenders with a valuable method of
detecting malicious domains with DNS. The difference in capabilities between
the attacker and defender can result in the defender collecting more benign
data and training a cGAN with a better-performing discriminator.

\paragraph{Mitsuhashi et al. 2023} propose a system to detect DGA-based malware
communications from DNS over HTTPS (DoH) traffic~\cite{mitsuhashi2023}.  The
DoH protocol is designed to protect the privacy of Internet users, but it can
also prevent network administrators from detecting suspicious communications.
The proposed system uses hierarchical ML analysis to classify network traffic
with gradient boosting decision tree and tree-ensemble models. The evaluation
shows that the system can detect DoH traffic generated by different types of
DGA-based malware with \textbf{99.12\%} accuracy, which can support network
administrators in improving anti-malware solutions.

\section{Discussion} \label{sec:disc}
In this section we highlight the results from the literature study and identify
trends in the areas of RPZ, TIF, and DGA detection.  We also analyze the
intersections in the literature between the various techniques, discuss the
misuse of filtering DNS resolvers and present future research grounded in our
findings.

\subsection{Trends}
\paragraph{RPZ} (Response Policy Zones) has been recognized as a potent DNS
firewall for filtering out malicious domains and URLs, notably those embedded
in emails~\cite{jin2020, jin2023}. However, selecting appropriate block/allow
lists poses a critical challenge to ensure that legitimate domains are not
inadvertently blocked.  Static lists may fall short in addressing the rapidly
evolving threat landscape, necessitating the integration of up-to-date feeds
into RPZ to detect emerging threats, such as those identified by TIF using DGA
detection.  Effective clustering of malicious domains may be essential to
streamline investigations and prevent overwhelming analysts during the
investigative process~\cite{satoh2019}. A study has also delved into whether
end-users should have the capability to interact with the blocking process,
highlighting the importance of involving end-users in the overall security
mechanism~\cite{jin2021}.

The challenge of selecting appropriate block/allow lists for RPZ cannot be
overstated. Striking the right balance to prevent blocking legitimate domains
while effectively identifying malicious ones is a constant concern. This
challenge is exacerbated by the dynamic nature of threats, making static lists
insufficient for robust defense. Real-time threat intelligence integration is
imperative for RPZ to promptly adapt and respond to emerging threats. Moreover,
without efficient clustering methods to group related malicious domains, the
task of investigating and mitigating threats becomes overwhelming for analysts.
Therefore, investing in advanced clustering techniques is crucial to empower
analysts and enhance the efficacy of the RPZ system.

\paragraph{TIF} (Treat Intelligence Feeds) is a valuable technique for
enhancing DNS filtering. However, significant challenges need to be addressed
to maximize its effectiveness. These challenges pertain to feed latency, feed
quality, and the integrity of the training data. Timely and accurate threat
intelligence is crucial for minimizing false positives in client-side
detection~\cite{arnaldo2018}. Additionally, training data from popularity lists
is vulnerable to manipulation, highlighting a key integrity
concern~\cite{rweyemamu2019}. Studies emphasize the importance of enhancing the
overall quality of TIF through measures such as adopting privacy-preserving
intelligence sharing among organizations~\cite{preuveneers2022} and providing
more structured information about the feeds to customers~\cite{li2019a,
griffioen2020}.

One prominent challenge associated with TIF is providing low latency and high
quality in threat feeds, ensuring the timely detection and blocking of
malicious domains, thus preventing potentially irreversible damage. Equally
significant is the challenge of maintaining the integrity of training data.
False positives and negatives must be minimized, especially when utilizing
popularity lists as a basis for training. Furthermore, providing customers with
structured information about the feeds is essential for enabling better
comprehension and effective utilization of the threat intelligence provided.

\paragraph{DGA} (Domain Generation Algorithms) is a tool frequently leveraged
by malware to avoid detection by generating dynamic domain names. Detection
techniques using ML have been employed to counter this evasion strategy. The
current research landscape underscores a growing trend in utilizing DL and
GANs~\cite{arnaldo2018, li2019b, satoh2021, mohus2022}, as well as explainable
AI (XAI)~\cite{drichel2021} in this domain. However, detecting unknown and
novel DGAs poses a substantial challenge, necessitating the continuous
accumulation of additional contextual information.

When evaluating the outcomes documented in the existing literature on DGA
detection, promising results are evident, with reported accuracies often
reaching 99\% and F1-scores achieving 0.80. Nevertheless, a critical
consideration revolves around the relatively small portion of network traffic
that constitutes algorithmically generated domains, also known as the ``base
rate''. This element presents a significant challenge in effectively reducing
false positives, even when detection models exhibit high reported accuracies.
Furthermore, it is imperative to explore cases where DNS traffic mirrors actual
communication from infected hosts and devise methods to identify instances of
off-use DNS resolutions~\cite{wala2022}.

Detecting unknown and novel DGAs presents an ongoing challenge, necessitating
the perpetual accumulation of contextual information and the flexibility to
adapt to evolving algorithms. Despite seemingly high accuracy in detection
models, the persistent issue of minimizing false positives underlines the need
for nuanced approaches to DGA detection. These nuances have far-reaching
implications for cybersecurity, emphasizing the importance of continuously
refining detection methodologies to effectively keep pace with evolving malware
tactics and minimize the risk of false positives.

\subsection{Intersections}
\begin{figure}
    \centering
    \includegraphics[trim={27cm 0 0 57cm},clip, width=.6\textwidth]{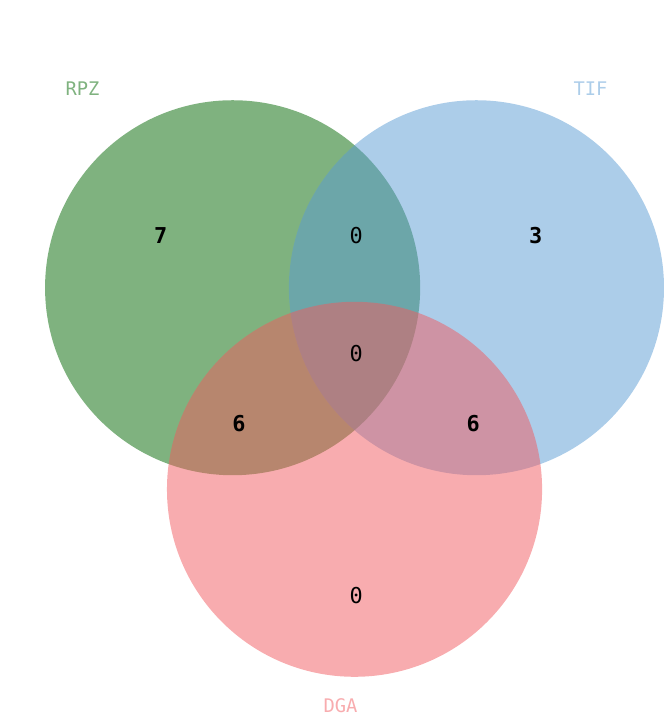}
    \caption{Intersection of topics}
    \label{fig:venn}
\end{figure}
Given the scope of this study, during 2018 and 2019, there was a significant
focus on Threat Intelligence Feeds in the literature, with a notable amount of
studies dedicated to exploring their capabilities, quality, and efficacy in
threat detection and prevention. However, starting from 2020 and continuing
into subsequent years, there has been a notable shift toward Response Policy
Zones (see Table~\ref{table:papertopic}), addressing their potential and
application in filtering DNS traffic.

Filtering resolvers can be enhanced with the intersection of different
techniques, such as RPZ, TIF, and DGA detection. However, there are missing
intersections in the literature that could improve their effectiveness (see
Figure~\ref{fig:venn}). For instance, RPZ and TIF can complement each other by
effectively distributing and sharing intelligence between them, such as
blocklists and allowlists. Moreover, DGA detection could enhance capabilities
of TIF, identifying and classifying malicious domains that evade
traditional methods. 

\subsection{DNS Filtering Misuse}
Employing DNS filtering to block or redirect oblivious users to alternative
services could be a censorship tool, reinforcing efforts to divert users away
from content that challenges prevailing narratives~\cite{hoang2021}. This
practice raises concerns about the suppression of free expression and the
potential for misuse in censoring information. Furthermore, such redirection
could risk users' sensitive data and personal information if directed to
potentially malicious websites or services. This dual impact on censorship and
security underscores the critical need for responsible and transparent DNS
filtering practices.  It is, therefore, imperative to establish a transparent
system that users can \emph{trust}. The filtering process should be based on
clear and objective criteria, with publicly available rules, and subject to
regular review and scrutiny. Users should have the option to opt out of the
filtering process, and an independent oversight mechanism should ensure the
integrity and impartiality of filtering practices.

\subsection{Proposed Future Work on Framework Design}
We propose future work on designing an open-source framework using RPZ, TIF,
and DGA detection to improve the filtering of malicious domains.  RPZ would be
used to intercept queries and match patterns against policies using blocklists
and allowlists, keeping the lists up-to-date using TIF.  The TIF in turn would
utilize state-of-the-art DGA detection techniques on anonymized resolver
queries to identify and classify known and unknown malicious domains.  The
system would use XAI and privacy-preserving intelligence sharing between
organizations to enhance the quality of the feeds and provide customers with
more structured information about the feeds. This framework would provide a
comprehensive solution for detecting and blocking malicious domains while
minimizing false positives and ensuring timely and accurate threat
intelligence.  In this framework, users would be able to see the filtering
process by making the filter lists publicly available and understanding blocked
domains.  This transparency helps build trust and ensures the filtering system
is not being misused for censorship. When the filter lists are accessible to
the broader security community, researchers, network administrators, and other
stakeholders can contribute their expertise and insights.  The argument against
publishing the filtered information concerns that the filtering system would
leak information to malware creators. However, as the filters become known,
they can force the malware creators to change their tactics, potentially making
them more visible in network traffic and easier to identify. This change, in
turn, can help security researchers and defenders gain insights into new
evasion techniques employed by malware creators.

The RPZ component filters DNS queries based on predefined policies using
blocklists and allowlists.  The resolver implementing the RPZ component may
also be responsible for sharing privacy-preserved traffic data with the threat
intelligence.  The threat intelligence analyzes the contextual patterns from
received traffic data, such as frequency and return codes. It may also
implement DGA detection to find patterns in potentially generated domains often
used by malware, providing an additional layer of sophistication in identifying
potential threats. The TIF component continuously enriches RPZ's policies by
updating the blocklists and allowlists with the latest known malicious domains
and IP addresses, closing the loop in the ecosystem.  By only sending
aggregated traffic data from the resolver to the threat intelligence we aim to
preserve the privacy of the resolver's users.  Our plan to optimize the
proposed system's security involves implementing robust access controls and
encryption methods to safeguard sensitive data and settings, thereby
discouraging unauthorized access and tampering. We will conduct thorough
security assessments, including penetration testing, to assess the framework's
resilience against various attack methods and potential misuse.  Our focus will
also be on enhancing accuracy and efficiency by rigorously validating and
verifying threat intelligence feeds, minimizing false positives and negatives
through a stringent validation process. Additionally, we intend to enhance
system availability and resilience by incorporating redundancy and failover
measures to mitigate unexpected failures or malicious attacks.

\section{Summary and Conclusion} \label{sec:con}
We performed a literature survey around filtering resolvers in DNS.  Focusing
on three techniques for implementing and enhancing DNS filtering we analysed
the trends since 2018 and observed partial intersections in the literature.
Surprisingly, no study covered both RPZ and TIF.  Combining these techniques
could improve their effectiveness in filtering malicious domains. However, each
area has its own challenges that need to be addressed.  Therefore, we propose
future research on designing an open-source framework that combines RPZ, TIF,
and DGA detection, addressing the gaps and challenges identified in the
literature.  

\bibliographystyle{plain}
\bibliography{ref} 

\end{document}